# Yttrium-substituted Mg-Zn ferrites: correlation of physical properties with Yttrium content

M.A. Ali[1], M.N.I. Khan[2], F.-U.-Z. Chowdhury[1], M.M. Hossain[1], A.K.M. Akhter Hossain[3], R. Rashid[2], A. Nahar[2], S.M. Hoque[2], M.A. Matin[4] and M.M. Uddin[1*]

[1]Department of Physics, Chittagong University of Engineering and Technology (CUET), Chattogram 4349, Bangladesh
[2]Materials Science Division, Atomic Energy Center, Dhaka 1000, Bangladesh.
[3]Department of Physics, Bangladesh University of Engineering and Technology (BUET), Dhaka 1000, Bangladesh.
[4]Department of Glass and Ceramic Engineering, Bangladesh University of Engineering and Technology (BUET), Dhaka 1000, Bangladesh.

**ABSTRACT**

Yttrium- (Y) substituted Mg-Zn ferrites with the compositions of $Mg_{0.5}Zn_{0.5}Y_xFe_{2-x}O_4$ ($0 \leq x \leq 0.05$) have been synthesized by conventional standard ceramic technique. The effect of $Y^{3+}$ substitution on the structural, electrical, dielectric and magnetic properties of Mg-Zn ferrites has been studied. The single phase of spinel structure with a very tiny secondary phase of $YFeO_3$ for higher Y contents has been detected. The theoretically estimated lattice constant has been compared with measured experimental lattice constant. The bulk density, X-ray density and porosity have been calculated. The Energy Dispersive X-ray Spectroscopy (EDS) study confirms the presence of Mg, Zn, Y, Fe and O ions in the prepared samples. Frequency dependence of conductivity has been studied and an increase in resistivity (an order) has been observed. Frequency dependence of dielectric constant ($\varepsilon'$), dielectric loss tangent ($\tan\delta$) has been studied and the lowering of $\varepsilon'$ with the increase of Y content was noted. Dielectric relaxation time was found to vary between 15 to 31 nano seconds. The saturation magnetization ($M_s$), coercive field ($H_c$), remanent magnetization ($M_r$) and Bohr magneton ($\mu_B$) have been calculated. The variation of $M_s$ has been successfully explained with the variation of A–B interaction strength due to Y substitution. The soft ferromagnetic nature also confirmed from the values of $H_c$. The complex permeability has been studied and the initial permeability was found to increase with Y up to $x = 0.01$, thereafter it decreases. The values of electrical resistivity and dielectric constant with proper magnetic properties suggest the suitability of Y-substituted Mg-Zn ferrites in microwave device applications.

**Keywords:** Mg-Zn ferrites, structural properties, cation distribution, electrical and dielectric properties, magnetic properties and permeability.

*Corresponding author: mohi@cuet.ac.bd

## 1 Introduction

The field of ferrites has become a possible subfield of materials science because of unique combination of their electrical and magnetic properties. The physics involved in ferrites have also drawn much attention to the scientific community and interests in ferrites are still now growing, even after many years since their discovery. The scientists, researchers, technologist and engineers are continuously trying to open the door for commercial application of ferrites and various types of ferrites with excellent properties are known in bulk, thin film and nano particle form [1]. Among the large family of ferrites, the Mg-Zn ferrites with spinel structure are widely known due to their significant properties which makes them suitable for application in computer memory and logic devices, cores of transformers, recording heads, antenna rods, loading coils and microwave frequency devices (as a core of coils), and so forth [2, 3]. Like other soft ferrites, they are also considered as a suitable choice owing to their high Curie temperature ($T_c$), high electrical resistivity, low eddy current losses, low dielectric loss, low cost, high mechanical hardness and superior environmental stability [4, 5]. Even their application extends in every sector: electronic communication to medical industry, military to space technology [3].

The synthesis of new ferrites with different compositions with better performance in practical application by the modification of existing materials is always motivating the materials researchers [6]. The properties of ferrites strongly depend on the chemical composition, cation distribution on A-site and B-site, methods of preparation, sintering temperature and time, types of impurity ions and levels etc. [7, 8]. Therefore, there is a way of tuning the physical properties of ferrites by changing the chemical composition. The cation distribution at A-site and B-site can be changed by different ions substitution for either divalent cation ($M^{2+}$ of $MFe_2O_4$) or $Fe^{3+}$ ions. The aim of this research work is to enhance the electrical resistivity of Mg-Zn ferrites by substituting $Fe^{3+}$ with rare-earth ($Y^{3+}$) ions with an intension to achieve desired characteristics of ferrites for applications in the high frequency devices. The electrical conductivity in spinel ferrites is mainly due to the electron hopping between $Fe^{2+}$ (which are form during sintering process [9, 10]) and $Fe^{3+}$ ions at octahedral (B-site) site [11]. It is well known that the rare-earth ions occupy the octahedral (B) sites that reduced the electron hopping by limiting the motion of $Fe^{2+}$, results an increase in resistivity [12]. Therefore, increase of electrical resistivity is expected by $Y^{3+}$ substitution for $Fe^{3+}$ in Mg-Zn ferrites due to its ($Y^{3+}$) tendency to occupy the B-site due to larger ionic radius [13]. It is noted that the Yttrium (Y) has already been substituted in Ni-Zn ferrites, Mg ferrites and Co ferrites [13-20]. Enhancement of electrical resistivity in Y substituted Ni-Zn and Mg ferrites have also been reported [16-18].

The physical properties of different ions substituted Mg-Zn have been investigated by many researchers: Mn [2, 3, 21], Sm [11], Ti [22], Nd [23, 24], Zr [25], Cu [26-29], Cr [30-32], Co [33, 34], Pr [35], Gd [36], Ni [37, 38] of which some of them were dealt with electrical and dielectric properties [3, 11, 21, 22, 26, 29, 32, 38] and some of them were dealt with magnetic properties [2, 25, 28, 31, 34-37] of substituted Mg-Zn ferrites. Few of them were also dealt with both of dielectric and magnetic properties [27, 30, 33]. Improvement of electrical properties of Mg-Zn ferrites has also been reported by Nd [24, 25] and Sm [11] substitutions that have been explained by their larger ionic radius and B-site occupation. However, to the best of our knowledge, study of Y substituted Mg-Zn ferrites is not reported yet.

Therefore, the study of the structural, electrical, dielectric and magnetic properties of Mg-Zn ferrites as a function of Y contents has been done for the first time.

## 2 Experimental techniques

Conventional ceramic technique was used to prepare Y-substituted Mg-Zn ferrite [$Mg_{0.5}Zn_{0.5}Y_xFe_{2-x}O_4 (0 \leq x \leq 0.05)$]. The raw materials in nano form were used (US Research Nanomaterials, Inc.) to facilitate the homogeneous mixing with purity > 99.5% and the particle size are of 20, 10-30, 30 and 20-40 nm for MgO, ZnO, $Fe_2O_3$ and $Y_2O_3$, respectively. We have already completed some projects using nano powders as raw materials and the results are available elsewhere [39-42]. The raw powders were weighed according to the stoichiometric ratio for corresponding composition. The powders were then mixed and milled for 6hrs using an agate mortar and pestle. After completing the homogeneous mixing of powders, they were loosely pressed to make biscuit like shape and then were calcined at 850 °C for 4 hrs in a muffle furnace. The calcined powders were then milled again for 2.5 hrs. A 5% polyvinyl alcohol solution was added as a binder and desired shape of dimension 8.4 mm diameter and 2.4 mm thickness of samples were prepared using a suitable die with a hydraulic press by applying 10 kN pressure. The samples were finally sintered at 1250 °C for 4 hrs in air at atmospheric pressure with the temperature step of 5 °C per minute and cooled naturally. The characterization of samples was done by taking X-ray diffraction (XRD) using Philips X'pert PRO X-ray diffractometer (PW3040) with Cu-K$_\alpha$ radiation ($\lambda$=1.5405 Å), microstructure images with the EDS by a high resolution FESEM (JEOL JSM-7600F), dielectric and permeability measurements by a Wayne Kerr precision impedance analyzer (6500B) in the frequency range of 10–120 MHz with a drive voltage of 0.5 V at room temperature. The room temperature magnetic properties were obtained by a physical properties measurement system (PPMS) from quantum design.

## 3 Results and discussion

### 3.1 Structural properties

The X-ray diffraction (XRD) patterns of Y-substituted Mg-Zn ferrites with the chemical composition, $Mg_{0.5}Zn_{0.5}Y_xFe_{2-x}O_4$ ($0 \leq x \leq 0.05$) are shown in Fig. 1. The sharp and well defined peaks in the XRD patterns [ICDD PDF 22-1012] confirmed the spinel structure in which peaks are assigned for Miller indices. The single phase of spinel structure is identified up to $x$=0.03, there after a secondary extra phase is observed at $2\theta = 33.2°$. The secondary phase appears at the grain boundaries (marked as red filled circles in Fig. 1) results from $YFeO_3$ [ICDD PDF # 39-1489] due to high reactivity of $Fe^{3+}$ ions with $Y^{3+}$ ions [43].

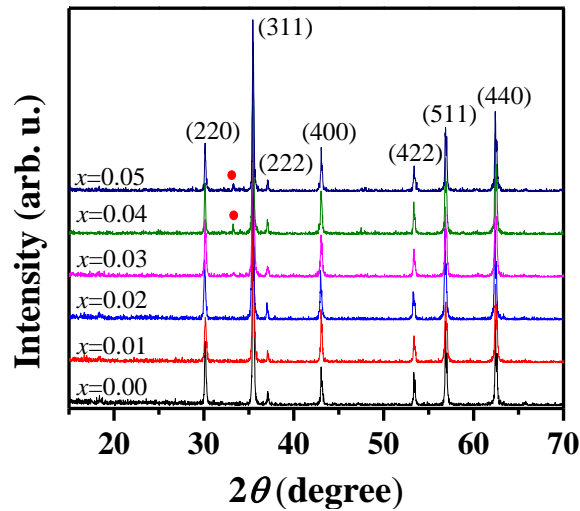

**Fig. 1** The XRD patterns of $Mg_{0.5}Zn_{0.5}Y_xFe_{2-x}O_4$ ($0 \leq x \leq 0.05$).

The similar phase is also reported in Y-substituted Ni-Zn ferrites [16] and Mg ferrites [18]. The XRD data has also been used to determine the lattice constants of the samples.

To calculate the lattice constant (*a*), the formula has been used for all peaks of the samples is given by: $a = d\sqrt{h^2 + k^2 + l^2}$, where *h*, *k* and *l* are the Miller indices of the crystal planes. The Nelson-Riley (N-R) extrapolation method has been used to evaluate lattice constants for the samples, the N-R function, $F(\theta)$, is given as $F(\theta) = \frac{1}{2}\left[\frac{cos^2\theta}{sin\theta} + \frac{cos^2\theta}{\theta}\right]$ [44]. The values *a* for each peaks are plotted against $F(\theta)$ and the exact lattice constant has been obtained from the point where the least square fit straight line cut the y-axis.

The theoretical density also known as X-ray density has been calculated using following expression: $\rho_x = \frac{8M}{N_A a_0^3} g/cm^3$, where $N_A$ is Avogadro's number (6.02×10²³ mol⁻¹), *M* is the molecular weight. The bulk density is measured by the formula: $\rho_b = \frac{M}{V} g/cm^3$ where $V (=\pi r^2 h)$ is the volume of the samples, *r* and *h* are radius and height of the samples. Porosity (P) of the samples is calculated using the following equation: $P = \left(\frac{\rho_x - \rho_b}{\rho_x}\right) \times 100\%$. where, $\rho_b$ is the bulk density of the samples and $\rho_x$ is the X-ray density. The calculated lattice constant, bulk density, X-ray density, and porosity are presented in Table 1.

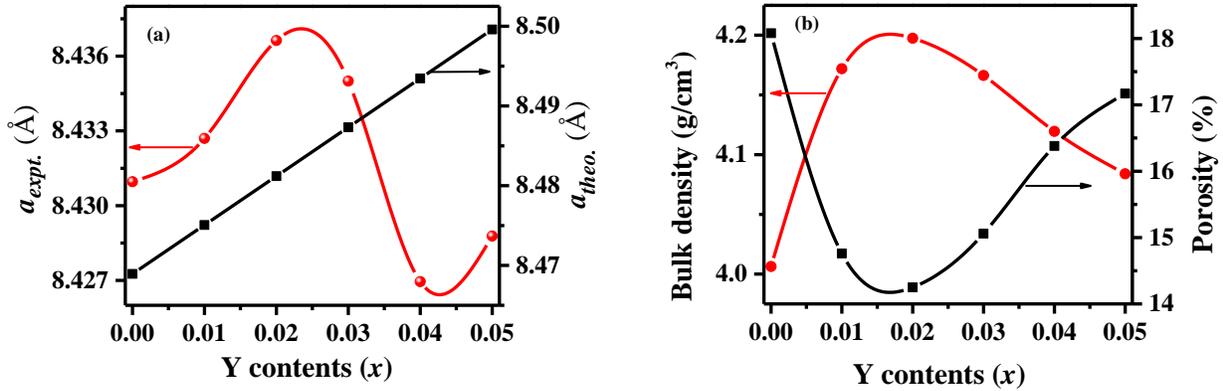

**Fig. 2** (a) The obtained lattice constants (experimental and theoretical) and (b) bulk density and porosity of $Mg_{0.5}Zn_{0.5}Y_xFe_{2-x}O_4$ ($0 \leq x \leq 0.05$).

Fig. 2 (a) illustrates the calculated lattice constants (experimental and theoretical) as a function of Y contents. Increase in lattice constants up to *x* = 0.03 (expt.) are due to substitution of larger ionic radius of Y ion (0.9 Å) compared to Fe ion (0.67 Å). On the other hand, the values of *a* are noted to be decreased for higher Y contents. An extra impurity phase is also observed (Fig. 1) for $x = 0.04$ and $0.05$, indicating diffusion of some of Y ions into the grain boundaries consequently the $YFeO_3$ is formed. The spinel lattice is compressed due to the differences in the thermal expansion coefficient in presence of extra inter-granular phase, results a decreasing trend in $a$ [12] that has been reported by others [16-18]. Fig. 2 (a) also depicts the theoretical lattice constant ($a_{theo}$), calculated using the equation which relates lattice constant with the ionic radii of $A$ and $B$ crystallographic lattice sites [45]: $a_{th} = \frac{8}{3\sqrt{3}}[(r_A + R_0) + \sqrt{3}(r_B + R_0)]$, where $R_0$ is the radius of the oxygen ion (1.32 Å) [46]. The ionic radii of $r_A$ and $r_B$ have been calculated by assuming possible cation distribution. The cation distribution has been proposed based on the following assumptions: the $Mg^{2+}$ ions occupy both the $A$-sites (10 % of total distribution) and $B$-sites (90% of total distribution) [47] while $Zn^{2+}$ prefers to occupy the tetrahedral sites ($A$-sites) [48]. The $Fe^{3+}$ ions have preference for both tetrahedral and octahedral sites [49]. The mean ionic radii $r_A$ of tetrahedral sites (A) and $r_B$ of octahedral sites (B) have been calculated using relations [45];

$r_A = C_{AFe}r(Fe^{3+}) + C_{AMg}r(Mg^{2+}) + C_{AZn}r(Zn^{2+})$ and $r_B = C_{BFe}r(Fe^{3+}) + C_{BMg}r(Mg^{2+}) + C_{BY}r(Y^{3+})$. A good agreement is to be noted between theoretical and experimental lattice constant up to $x = 0.03$. The theoretical lattice constant ($a_{theo}$) is found to be higher than the experimental lattice constant ($a_{expt}$). In order to calculate the theoretical lattice constant the ideal crystal unit cell is considered. The ideal unit cell is perfectly filled in a close packed spinel structure manner with regular cation and anions distribution. Therefore, a little deviation between $a_{expt}$ and $a_{theo}$ is expected [50]. The $a_{expt}$ values are decreasing trend for $x = 0.04$ and $0.05$ following disagreement between $a_{expt}$ and $a_{theo}$ since the secondary phase is not considered in calculation of $a_{theo}$, however it is obvious for $x = 0.04, 0.05$ and seems to be started at 0.03 (Fig. 1).

For different Y contents, the values of X-ray density ($\rho_x$) and bulk density ($\rho_b$) are presented in Table 1. Pores formed during sintering process are not considered in calculation of $\rho_x$ that produces higher values of $\rho_x$ than that of $\rho_b$. Fig. 2(b) represents the variation of $\rho_b$ and porosity ($P$) with Y contents. The value of $\rho_b$ increases up to $x = 0.02$ and beyond this it decreases. The appearance of impurity phase ($YFeO_3$) in Fig. 1 can be explained in a way that the inter-granular voids might not be filled for higher Y contents. It should be noted here that the values of $\rho_b$ for substituted compositions are higher than that of the un-substituted one due to the difference in atomic weight of Y (88.90585 amu) and Fe (55.845 amu). The variation of density with Y contents can also be more cleared in the subsection 3.2.

The *A*-sites are small in volume due to movement of oxygen ions to adjust the metal ions consequently A-sites expand to an extent corresponding to the reduction of B-sites [49]. The oxygen positional parameter (*u*), along with the bond lengths at tetrahedral sites ($R_A$) at octahedral sites ($R_B$), the tetrahedral edge length R, shared octahedral edge length R' and unshared octahedral edge length R'' have been calculated (Table 1) using following equations:

$$u = \left[\frac{1}{a_{th}\sqrt{3}}(r_A + R_0) + \frac{1}{4}\right][45], R_A = a\sqrt{3}\left(\delta + \frac{1}{8}\right), R_B = a\left(\frac{1}{16} - \frac{\delta}{2} + 3\delta^2\right)^{1/2}[51],$$

$$R = a\sqrt{2}(2u - 0.5), R' = a\sqrt{2}(1 - 2u) \text{ and } R'' = a\sqrt{4u^2 - 3u + {}^{11}/_{16}} \text{ [45]},$$ where $\delta$ (= $u - u_{ideal}$), is the inversion parameter which signifies departure from ideal oxygen parameter ($u_{ideal}$=0.375 Å) and *a* is the experimental lattice constant.

The bond length $R_A$ and non-linear behavior of the bond length $R_B$ with Y substitution are observed and shown in Table 1. The cation distribution on *A*-sites and *B*-sites results the variation of the tetrahedral edge length R, shared octahedral edge length R' and unshared octahedral edge length R'' [52]. The R is found to decrease with Y substitution. On the other hand, the R' and R'' are found to increase with Y contents. The calculated results are in good agreement with that of the Ni substituted Mg-Zn ferrites [37].

**Table 1** Cation distribution (A- and B-sites), ionic radii ($r_A$ and $r_B$), theoretical ($a_{theo.}$) and experimental ($a_{expt.}$) lattice constants, X-ray density ($\rho_x$), bulk density ($\rho_b$), porosity ($P$), bond length $R_A$ and $R_B$, tetrahedral edge length ($R$), shared octahedral edge length ($R'$) and unshared octahedral edge length ($R''$) of $Mg_{0.5}Zn_{0.5}Y_xFe_{2-x}O_4$ ($0 \leq x \leq 0.05$) with Y contents.

| Y contents (x) | A-site | B-site | $r_A$ (Å) | $r_B$ (Å) | $a_{theo}$ (Å) | $a_{expt.}$ (Å) | $\rho_x$ (g/cm³) | $\rho_b$ (g/cm³) | P (%) | u (Å) | $R_A$ (Å) | $R_B$ (Å) | R (Å) | R' (Å) | R'' (Å) |
|---|---|---|---|---|---|---|---|---|---|---|---|---|---|---|---|
| 0.00 | $Fe_{0.45}Mg_{0.05}Zn_{0.5}$ | $[Fe_{1.55}Mg_{0.45}]O_4^{2-}$ | 0.744 | 0.665 | 8.4689 | 8.4309 | 4.89 | 4.006 | 18.08 | 0.3907 | 2.0547 | 1.9841 | 3.3554 | 2.6178 | 3.0060 |
| 0.01 | $Fe_{0.45}Mg_{0.05}Zn_{0.5}$ | $[Fe_{1.54}Mg_{0.45}Y_{0.01}]O_4^{2-}$ | 0.744 | 0.666 | 8.4720 | 8.4327 | 4.894 | 4.172 | 14.76 | 0.3906 | 2.0544 | 1.9849 | 3.3549 | 2.6200 | 3.0070 |
| 0.02 | $Fe_{0.45}Mg_{0.05}Zn_{0.5}$ | $[Fe_{1.53}Mg_{0.45}Y_{0.02}]O_4^{2-}$ | 0.744 | 0.668 | 8.4750 | 8.4366 | 4.895 | 4.197 | 14.25 | 0.3905 | 2.0546 | 1.9862 | 3.3553 | 2.6221 | 3.0080 |
| 0.03 | $Fe_{0.45}Mg_{0.05}Zn_{0.5}$ | $[Fe_{1.52}Mg_{0.45}Y_{0.03}]O_4^{2-}$ | 0.744 | 0.669 | 8.4781 | 8.4350 | 4.905 | 4.166 | 15.06 | 0.3904 | 2.0534 | 1.9861 | 3.3534 | 2.6243 | 3.0090 |
| 0.04 | $Fe_{0.45}Mg_{0.05}Zn_{0.5}$ | $[Fe_{1.51}Mg_{0.45}Y_{0.04}]O_4^{2-}$ | 0.744 | 0.670 | 8.4812 | 8.4269 | 4.927 | 4.119 | 16.38 | 0.3903 | 2.0507 | 1.9846 | 3.3490 | 2.6265 | 3.0100 |
| 0.05 | $Fe_{0.45}Mg_{0.05}Zn_{0.5}$ | $[Fe_{1.50}Mg_{0.45}Y_{0.05}]O_4^{2-}$ | 0.744 | 0.671 | 8.4842 | 8.4288 | 4.931 | 4.084 | 17.17 | 0.3902 | 2.0505 | 1.9854 | 3.3485 | 2.6286 | 3.0110 |

**Table 2** Cations-anions atomic % and cations to anions ratio of $Mg_{0.5}Zn_{0.5}Y_xFe_{2-x}O_4$ ($0 \leq x \leq 0.05$).

| Y contents (x) | Grain size (μm) | cations (At %) | anions (At %) | Ratio of cation: anion |
|---|---|---|---|---|
| 0.00 | 1.22 | 40.03 | 59.97 | 2.802:4.197 |
| 0.01 | 3.65 | 40.68 | 59.32 | 2.847:4.152 |
| 0.02 | 3.75 | 40.29 | 59.71 | 2.820:4.179 |
| 0.03 | 1.28 | 42.09 | 57.91 | 2.946:4.053 |
| 0.04 | 1.07 | 47.94 | 53.94 | 3.355:3.775 |
| 0.05 | 1.87 | 38.40 | 61.60 | 2.688:4.312 |

## 3.2 Microstructure study

The role of microstructure in materials designing is very important to obtain the desired properties for their useful applications. The physical properties can significantly be tuned by changing the microstructure of materials. Fig. 3 demonstrates the microstructure of $Mg_{0.5}Zn_{0.5}Y_xFe_{2-x}O_4$ ($0 \leq x \leq 0.05$). A significant effect of Y ions substitution on the microstructure is clearly identified from the images of Fig. 3.

Significant change in shape and size of grains and grain boundaries is observed. Moreover, the densification of Y substituted samples can also be observed from the figure. An inverse relation is found between the variation of bulk density and porosity [Fig. 2 (b)] and grains size distribution as function of Y contents. The increase in grains size with Y contents up to $x = 0.02$, indicating the appreciable extent of diffusion of Y ions into the Mg-Zn ferrites grains and most denser composition. After $x = 0.02$, the extent of diffusion decreases due to larger ionic radius of Y, residual stress is created, which results smaller grains and lowered the density. The variation of grain sizes is also associated with the variation of lattice constants with Y contents.

The elements of prepared samples can be detected using EDS. It is also possible to detect the existence of unwanted elements. The peaks and height of peaks in EDS spectra correspond to the elements present and concentrations of element. Each peak is unique for every element. The EDS spectra (not shown here) of the samples were obtained and the presence of Mg, Zn, Fe, Y and O elements are confirmed. The standard ratio of metal cations to anion in ferrites is 3:4 [37]. The obtained atomic % of metal cations and anions are presented in Table 2. A fairly good agreement with standard ratio is observed and deviation from standard ratio is also observed for Ni-substituted Mg-Zn ferrites [37]. The results obtained from XRD and FESEM measurements revealed the successful synthesis of Y-substituted Mg-Zn ferrites. Now, to explore the suitability the prepared samples the electrical, dielectric and magnetic properties of $Mg_{0.5}Zn_{0.5}Y_xFe_{2-x}O_4$ ($0 \leq x \leq 0.05$) ferrites have been studied which are presented in the following sections.

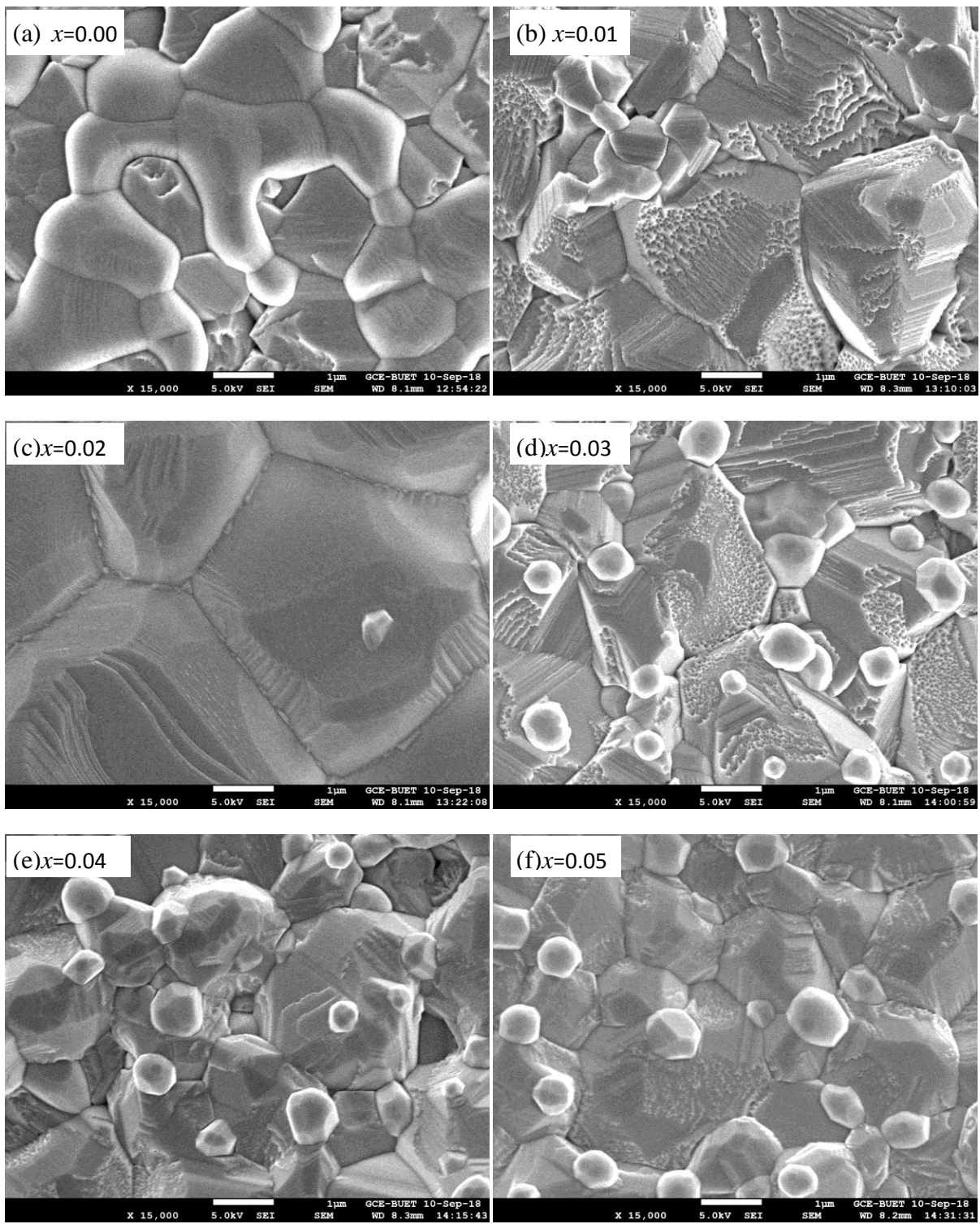

**Fig. 3** FESEM images of $Mg_{0.5}Zn_{0.5}Y_xFe_{2-x}O_4$ ($0 \leq x \leq 0.05$).

## 3.3 Frequency dependence of ac conductivity

Fig. 4 (a) shows the frequency dependence of ac conductivity ($\sigma_{ac}$) of the samples. The $\sigma_{ac}$ gradually increases with the increase in frequency. The frequency dependent total ac conductivity follows Jonscher's power law: [53]

$$\sigma_{ac,total}(\omega) = \sigma(0) + \sigma_{ac}(\omega) = \sigma_{dc} + A\omega^n$$

where, the first part ($\sigma_{dc}$) indicates the frequency-independent conductivity or dc conductivity. The pre-exponential factor 'A' is constant and dependent on temperature, 'n' is an exponent, dependent on both frequency and temperature in the range 0 to 1 and $\omega$ represents the angular frequency. Frequency dependent behavior of these samples can be explained by employing Koop's heterogeneous and Maxwell–Wanger double layer model [54-56]. The ferrites comprises of two layers: the layer consists of grains which are well conducting, separated by poorly conducting thin layer, forming grain boundary. The frequency response of these two layers is different resulting different conductivity at low and high frequency regions. At the lower frequency region, the dc conductivity is attributed due to the grain boundaries which are more active and the exchange of electrons between $Fe^{2+}$ and $Fe^{3+}$ ions is less. The grains activity increases with increasing the frequency by promoting the electrons hopping between $Fe^{2+}$ and $Fe^{3+}$ ions consequently an increase in hopping frequency hence ac electrical conductivity rises.

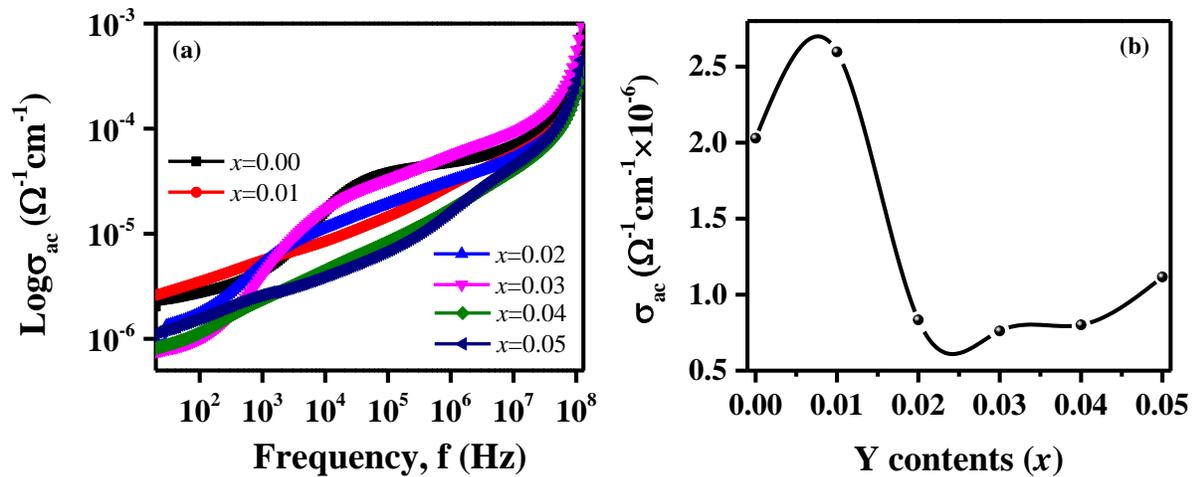

**Fig. 4** (a) The frequency dependence of ac conductivity for $Mg_{0.5}Zn_{0.5}Y_xFe_{2-x}O_4$ ($0 \leq x \leq 0.05$) and (b) ac conductivity for different Y contents at 20 Hz.

Fig. 4 (b) shows the variation of ac conductivity with Y contents at 20 Hz. The conductivity of Mg-Zn ferrites decreases with increasing Y contents unlike for $x = 0.01$ that is expected and consistent with reported Ho substituted Ni-Zn ferrites for $x = 0.015$ [57] and $x = 0.01$ [58]. When Y ions are added into Mg-Zn ferrites, the Y ions substitute the $Fe^{3+}$ ions at the B-site in the compositions [13]. At the same time, some of Zn ions loss compensated by pushing some $Fe^{3+}$ at the A-site. The electronic valence is higher for Fe ions than Zn ions; hence to balance the electrical charge, metallic vacancies which are charge carriers have been increased resulting increase in conductivity. The resistivity of the compositions rises with increasing Y contents that can be explained as: the $Fe^{3+}$ ions concentration at B-sites decrease due to increasing $Y^{3+}$ ions at B-sites that accelerates the conduction process between $Fe^{2+}$ and $Fe^{3+}$ at the B-site [59, 60] where $Fe^{2+}$ ions are produced during sintering process [30]. Therefore, the decrease in $Fe^{3+}$ concentration at B-site reduces the probability of electrons exchange between $Fe^{2+}$ and $Fe^{3+}$ and hence the resistivity is increased that consistent with reported results [18]. Another point could also be noted here that for $x = 0.04$ and 0.05, ac conductivity was noted to be increased but still lower than parent one. The excess of Y contents and $Fe^{2+}$ ions at the B-site might facilitate the conduction process [61]. The increase in the probability of electrons hopping is due to larger ionic radius of Y which causes the oxygen ions to close each other in the crystal lattice could be another probable reason. The similar result for La substituted Ni ferrites has also been reported [62].

**Table 3** The variation of resistivity and dielectric parameters with Y contents.

| Y contents ($x$) | $\rho_{ac}$ ($\times 10^5$) ($\Omega$-cm) | | $\varepsilon'$ ($\times 10^4$) | | $\varepsilon'\sqrt{\rho_{ac}}$ ($\times 10^6$) | | tan$\delta$ | $f_{max}$(tan$\delta$) ($\times 10^3$) | $\tau_{M''}$ (nanosec) |
|---|---|---|---|---|---|---|---|---|---|
| | at 20 Hz | at 1 kHz | at 20 Hz | at 1 kHz | at 20 Hz | at 1 kHz | at 20 Hz | | |
| 0.00 | 4.54 | 2.24 | 5.34 | 0.68 | 37.5 | 3.2 | 3.41 | 254.4 | 24.1 |
| 0.01 | 3.45 | 1.79 | 7.89 | 0.31 | 48.9 | 1.3 | 2.96 | 1.751 | 22.3 |
| 0.02 | 10.4 | 1.92 | 2.88 | 0.65 | 31.6 | 2.8 | 2.59 | 47.11 | 26.1 |
| 0.03 | 11.4 | 2.46 | 2.59 | 0.20 | 29.2 | 1.0 | 2.63 | 88.22 | 15.1 |
| 0.04 | 11.0 | 4.22 | 2.55 | 0.15 | 33.4 | 0.9 | 2.82 | 1.747 | 30.6 |
| 0.05 | 7.75 | 3.79 | 2.99 | 0.96 | 24.5 | 5.9 | 3.35 | 1.494 | 20.6 |

## 3.4 Dielectric properties

**Frequency dependence of dielectric constant**

The dielectric constant has been calculated using equation: $\varepsilon' = Cd/\varepsilon_0 A$, where $C$ is the capacitance, $d$ is the thickness, $\varepsilon_0$ is the permittivity in free space and $A$ is the surface area of pellet. The measured (at room temperature) frequency dependent dielectric constants $\varepsilon'$ of $Mg_{0.5}Zn_{0.5}Y_xFe_{2-x}O_4$ ($0 \leq x \leq 0.05$) are presented in Fig. 5 (a), showing a dispersion in $\varepsilon'$, to a different extent, in the low frequency range ($<10$ kHz) due to interfacial polarization.

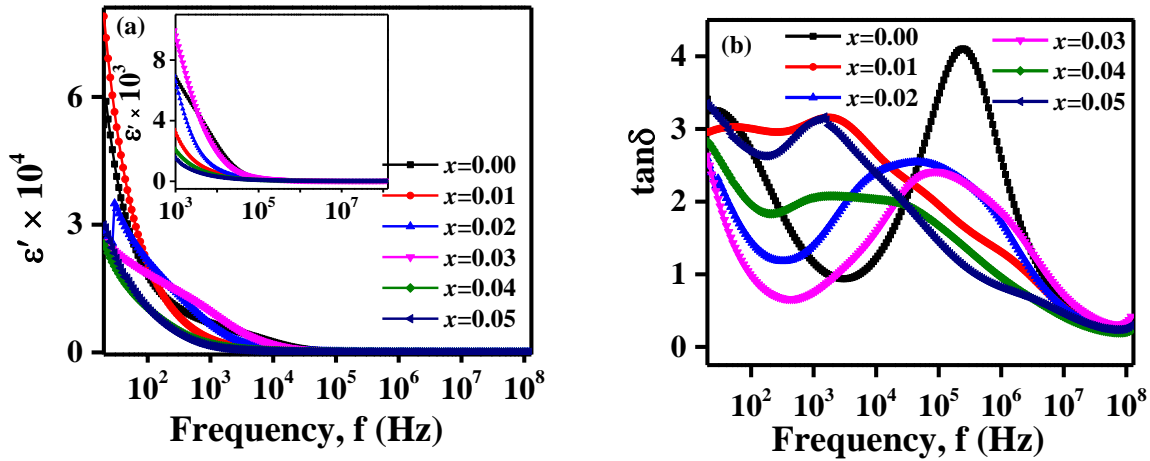

**Fig. 5** Frequency dependence of (a) dielectric constant and (b) dielectric loss tangent of $Mg_{0.5}Zn_{0.5}Y_xFe_{2-x}O_4$ ($0 \leq x \leq 0.05$). Inset of (a) shows the variation $\varepsilon'$ with frequency range from 1 kHz to 120 MHz.

The curves exhibiting three regions of frequencies: a sharp decrease up to 100 Hz, a slow decrease up to 10 kHz and finally become almost zero and frequency independent at high frequency region, which is common for spinel ferrites [63]. Similar results also reported earlier for Mg–Zn [64], Sm, Cr and Co substituted Mg-Zn ferrites [11, 32, 33] and Ni–Zn [40, 41]. The frequency dependent behavior of the compositions can be explained using of Koops' theory [54], assumes that the dielectric materials contain two layers of the Maxwell–Wagner type [55, 56]. The conduction mechanism and dielectric polarization is similar in ferrites [65] and a significant correlation between these two has also been reported [66]. The electron hopping between $Fe^{2+}$ and $Fe^{3+}$ has been taken in consideration for the polarization mechanism. In the low frequency region, the electron exchange between $Fe^{2+}$ and $Fe^{3+}$ is able to follow the electric field up to certain frequency (hopping frequency) results the local displacement of charges between sites in the applied field direction, which determine the polarization of the system. At high frequencies (after hopping frequency) the electron exchange between $Fe^{2+}$ and $Fe^{3+}$ cannot follow the applied electric field and hence attained a constant value [67]. It can also be noted here that the dielectric constant for $x = 0.01$ (Fig. 5 (a)) is lower than that of parent one but higher for other concentrations. The variation of dielectric constant of the compositions with Y contents can also be explained by employing same formalism as stated in subsection 3.3.

**Frequency dependence of dielectric loss**

An abnormal behavior is observed in the plots of the dielectric loss tangent (tanδ) against frequency for different Y contents, shown in Fig. 5 (b). The tanδ vs frequency curves for the sample showing typical maximum of dielectric loss at a certain frequency and the peak shift to lower frequency for substituted compositions. This type of behavior in Mg-Zn ferrites [68] as well as in Cu substituted Mg-Zn ferrites [27] has also been reported. This abnormal behavior is reported even in other ferrites such as Ni–Zn [41], Cu–Cd [69], Li–Mg–Ti [70] and Ni–Mg [71] ferrite systems. The condition for showing a peak in tanδ for a dielectric material can be expressed by the following relaxation relation $\omega\tau = 1$ [72], where $\omega = 2\pi f_{max}$ and $\tau$ is the relaxation time, related to the jumping probability per unit time $p$ by an equation $\tau = 1/2p$; or $f_{max} \propto p$ [27]. Therefore, the tanδ vs frequency curves exhibit a maximum when the jumps frequency between $Fe^{2+}$ and $Fe^{3+}$ ions at adjacent B-sites approximately equal to the applied electric field [73]. Thus the frequencies can be calculated (presented in Table 3) for samples at which the maximum in tanδ is observed.

**Correlation of resistivity and dielectric constants as a function of Y contents**

The conduction mechanism and dielectric polarization is similar in ferrites [65] and a significant correlation between these two has been reported [66]. In order to correlate the dielectric constant with ac resistivity, we have calculated the product of $\varepsilon'$ and $\sqrt{\rho_{ac}}$ (at 20 Hz and 1000 Hz) and presented in Table 3 from which an approximate inverse proportionality of $\varepsilon'$ to $\sqrt{\rho_{ac}}$ can be observed. Similar results have been reported and can be concluded that in case of dielectric materials where dielectric polarization is strongly dependent on conduction mechanism, the dielectric constant is approximately inversely proportional to the square root of resistivity [18, 74].

**Study of electric modulus**

Fig. 6 (a) shows the variation of real part of electric modulus of different compositions of $Mg_{0.5}Zn_{0.5}Y_xFe_{2-x}O_4$ ($0 \leq x \leq 0.05$). The value of M′ is very low (almost zero) in the frequency region around $< 10^4$ Hz and increases with increase in frequency showing a dispersion at frequency region $> 10^4$ Hz, exhibiting a maximum at very high frequency region and beyond it is down to a certain values. The characteristic feature might be demonstrated in the short range mobility of charge carriers due to the conduction process.

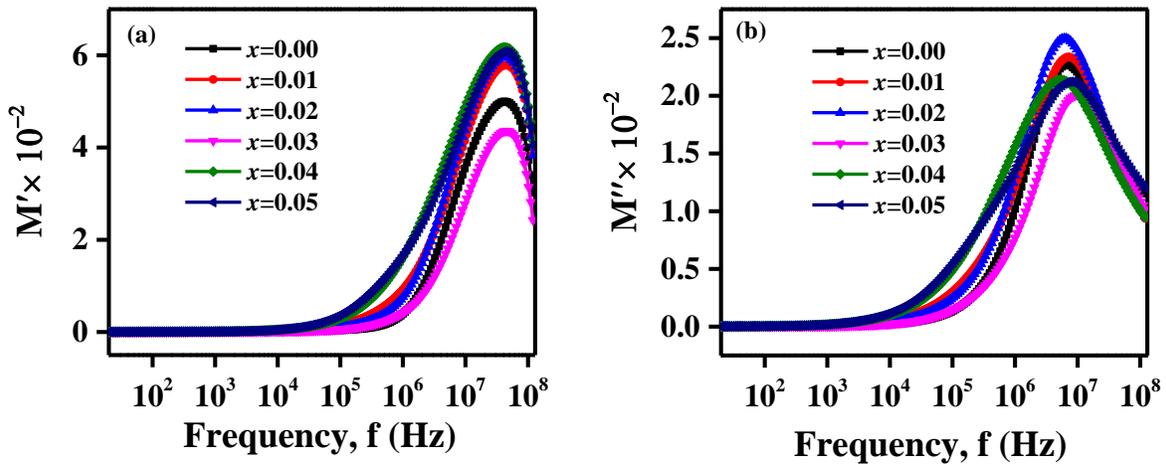

**Fig. 6** Frequency dependence of (a) real and (b) imaginary of electric modulus of $Mg_{0.5}Zn_{0.5}Y_xFe_{2-x}O_4$ ($0 \leq x \leq 0.05$).

Fig. 6 (b) shows the variation of imaginary electric modulus ($M''(\omega)$) of the samples with frequency, where the curves are characterized by (i) a clear peak in the pattern at different frequencies, (ii) The peaks are lying in the dispersion region of M′ and (iii) the peak shift towards lower frequency for $x = 0.02$ and 0.04 while the peak shifts towards high frequency side for $x = 0.01$, 0.03 and 0.05. The characteristic peaks distinguish the frequency range into two regions. The low frequency side that determines frequency range where charge carriers are able to move over a long distance i.e., hopping of charge carriers is possible between two adjacent sites. The high frequency region that determine frequency range where charge carriers are able to move within short range i.e., the motion of charge carriers are localized within their potential well. The peak also gives information about the transition from long range to short range mobility with increase in frequency [75]. Moreover, the frequency at which M″ is maximum is also known as dielectric relaxation frequency and the relaxation time τ is calculated using equation $\tau_{M''} = 1/2\pi f_{M''}$. The calculated relaxation time for different Y contents of the samples is presented in Table 3.

**Study of impedance spectroscopy**

Complex impedance is a useful tool to demonstrate the transport properties taking place between the grain and grain boundaries, resulting changes in electrical conductivity (ac and dc), dielectric permittivity, and dielectric losses. Especially, the dominant resistance from grains or grain boundary in the polycrystalline ceramics can easily be resolved [76]. Fig. 7 (a) and (b) represent frequency dependence of real and imaginary parts of the impedance for different Y contents. In Fig. 7 (a), the values of Z′ are found to decrease with increase in frequency, suggesting an increase in ac conductivity with frequency. The coincidence of the values Z′ at high frequency suggests the possible release of space charge [77] that is consistent with reported results [40, 41]. For particular frequency (e.g. at 20 Hz) the variation of Z′ with Y contents is same as observed and successfully discussed in case of conductivity and dielectric constants at same frequency. Fig. 7 (b) illustrates the imaginary part of the impedance Z″ as a function of frequency for different Y contents, showing a relaxation peak at low frequency side; all the curves coalesce with each other at high frequency.

The impedance plane plots of $Mg_{0.5}Zn_{0.5}Y_xFe_{2-x}O_4$ ($0 \leq x \leq 0.05$) compositions are presented in Fig. 8 (a). Usually, the impedance spectrum of the polycrystalline materials is characterized by the presence of one or more semicircles. The resistance of bulk grains or grain boundary can be calculated from the diameter of the semicircle at Z′ axis and the total resistance ($R_T = R_g + R_{gb}$) of the material is equal to intercept of the curve on the Z′ axis at low frequencies [78]. In case of ceramic materials, the contribution from grain boundary is much more due to existence of defects compared to the contribution from grain resistance. In Fig. 8 (a), two semicircles for each composition are observed but the semicircle at low frequency (right) remains incomplete within studied frequency range. Fig. 8 (b) shows the equivalent circuit model used to calculate the grain resistance ($R_1$), capacitance ($C_1$), grain boundary resistance ($R_2$) and capacitance ($C_2$); and the values are shown in Table 4. It is also clear from Table 4 that the grain resistance are too small compared to the grain boundary resistance for the compositions with $x$ = 0.00, 0.02, 0.03 and 05. On the other hand, the grain resistance are comparatively higher than grain boundary resistance in the composition of $x$ = 0.01 and 0.04. Similar results are also available for Gd substituted Ni ferrites [79]. It is assumed (for $x$ = 0.01 and 0.04) that the size of the grain boundaries is much thinner than bulk grains resulting lower values of grain boundary resistance compared to that of grain resistance [79]. To make a clear picture of the grains and grain boundaries, TEM/AFM or other very high resolution microscopy is necessary to produce the three dimensional image.

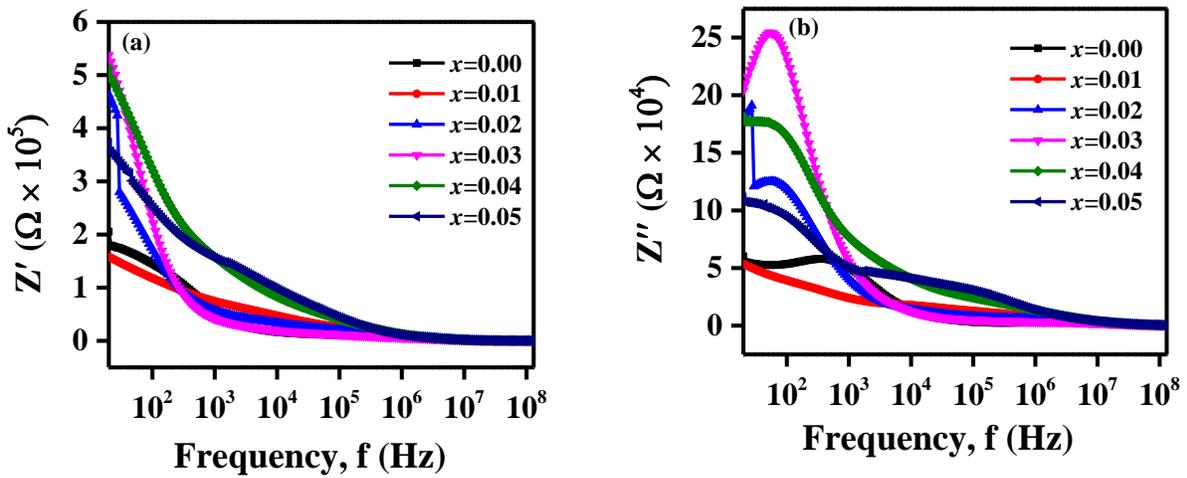

**Fig. 7** Frequency dependence of (a) real part and (b) imaginary part of impedance of $Mg_{0.5}Zn_{0.5}Y_xFe_{2-x}O_4$ ($0 \leq x \leq 0.05$).

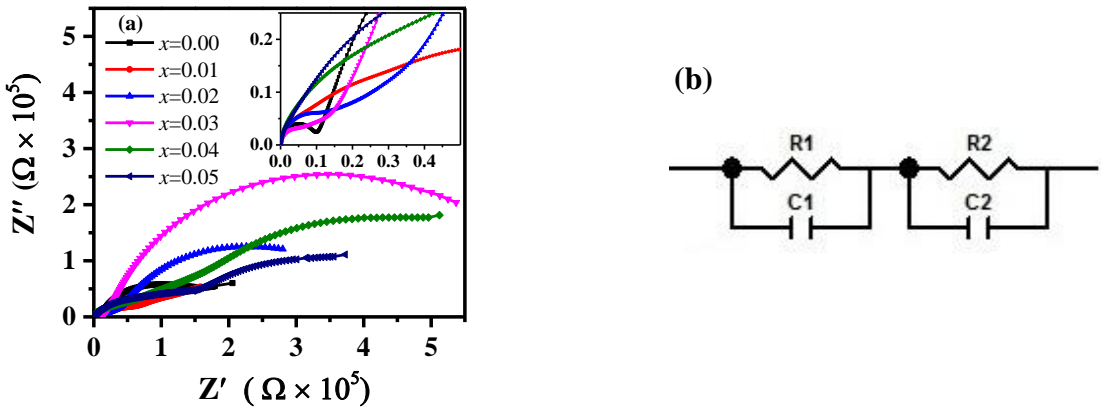

**Fig. 8** (a) The complex impedance spectra of $Mg_{0.5}Zn_{0.5}Y_xFe_{2-x}O_4$ ($0 \leq x \leq 0.05$) and (b) The equivalent model circuit for two semicircle Cole-Cole plot. The inset of (a) shows the first semicircle for $x$ = 0.00, $x$ = 0.02 and $x$ = 0.03.

**Table 4** The grain resistance ($R_1$) & capacitance ($C_1$) and grain boundary resistance ($R_2$) & capacitance ($C_2$).

| Y contents ($x$) | $R_1(\Omega)$ | $C_1(F)$ | $R_2(\Omega)$ | $C_2(F)$ |
|---|---|---|---|---|
| 0.00 | 1.1E4 | 8.1E-12 | 1.3E5 | 1.4E-9 |
| 0.01 | 6.7E4 | 7.8E-10 | 2.8E4 | 1.4E-11 |
| 0.02 | 2.3E4 | 1.3E-11 | 2.1E5 | 3.4E-9 |
| 0.03 | 1.3E4 | 2.1E-11 | 4.6E5 | 2.7E-9 |
| 0.04 | 2.7E5 | 1.0E-9 | 6.0E4 | 1.2E-11 |
| 0.05 | 6.4E4 | 1.1E-11 | 1.4E5 | 4.6E-10 |

### 3.5 Magnetic properties

The factors affecting the shape and the width of the hysteresis are the chemical composition of the compound, porosity, grain size, etc. The ferrimagnetic nature of all the samples has been confirmed from the field dependence of magnetization curve. Fig. 9 (a) shows narrow hysteresis loops confirming soft magnetic nature of all samples [80]. The prepared samples with very low coercivity (18-35 Oe) makes them suitable for use in high frequency devices and as core materials [81-84]. The saturation magnetization, coercive field, remanent magnetization and Bohr magneton have been calculated from the hysteresis curve and presented in Table 5.

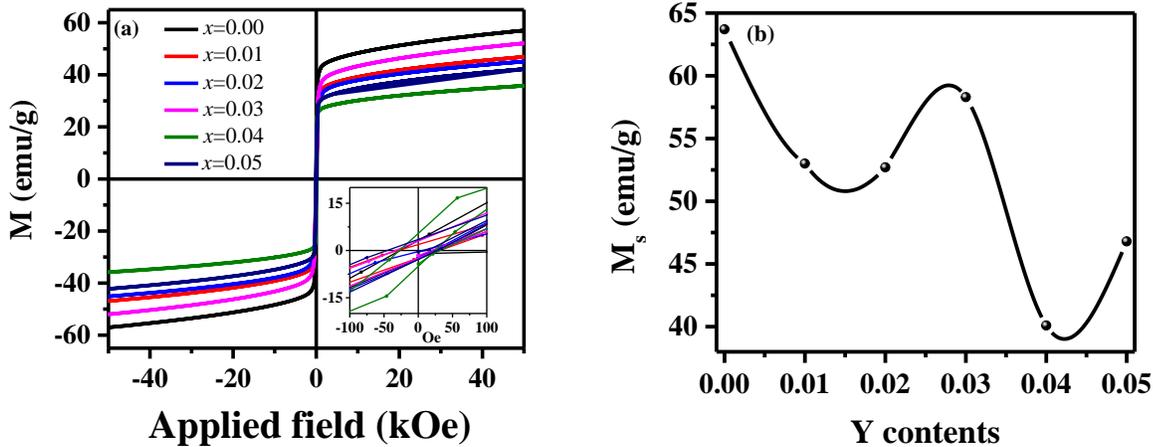

**Fig. 9** The (a) $M$–$H$ loops of $Mg_{0.5}Zn_{0.5}Y_xFe_{2-x}O_4$ ($0 \leq x \leq 0.05$) and (b) the variation of saturation magnetization with Y contents ($x$). The inset [Fig. 9 (a)] shows the hysteresis loops for very low applied field.

**Table 5** Saturation magnetization ($M_s$), coercive field ($H_c$), remanent magnetization ($M_r$) and Bohr magneton ($\mu_B$) of $Mg_{0.5}Zn_{0.5}Y_xFe_{2-x}O_4$ ($0 \leq x \leq 0.05$).

| Y contents ($x$) | $M_s$ (emu/g) | $M_r$ (emu/g) | $H_c$ (Oe) | $\mu_B$ |
|---|---|---|---|---|
| 0.00 | 63.7 | 2.55 | 26 | 2.51 |
| 0.01 | 53.0 | 2.38 | 30 | 2.08 |
| 0.02 | 50.1 | 2.84 | 18 | 1.99 |
| 0.03 | 58.3 | 2.43 | 31 | 2.31 |
| 0.04 | 40.1 | 5.04 | 26 | 1.59 |
| 0.05 | 46.8 | 2.84 | 35 | 1.86 |

From Fig. 9 (b) it is clear that the saturation magnetization for Y-substituted compositions are found to be less than the parent one ($x = 0.00$) due to the non-magnetic nature of Y ions. In this present case, non-magnetic $Y^{3+}$ is substituted for $Fe^{3+}$. But the decrease in $M_s$ is non-linear. The $M_s$ decreases up to $x = 0.02$ and then increase for $x = 0.03$, thereafter it decreases for $x = 0.04$ and finally slightly increases for $x = 0.05$. The variation of $M_s$ with Y contents can be explained by the exchange interactions of cations. Although AB interactions are strongest and dominant in the ferrites, but the AA and BB interactions also play an important role in determining the variation of magnetization. In Mg-Zn ferrites, the $Fe^{3+}$ ions occupy both A-site and B-site but the Y has a tendency to occupy the B-site only. When Y is substituted for $Fe^{3+}$ ions, it replaces some $Fe^{3+}$ from B-site only, results a decrease of magnetization at B sub-lattice. According to Neel two sub-lattice model the magnetization of ferrimagnetic is defined by the equation $M=M_B-M_A$, where $M_B$ and $M_A$ are magnetic moment of B sub-lattice and A sub-lattice. Therefore, the magnetization is expected to decrease for $Y^{3+}$ substituted compositions. The non-linear variation can also be explained based on the migration of $Mg^{2+}$ ions from B-sites to A-sites and cation redistribution. Generally, Mg ions occupy both the A-sites and B-sites. Some of Mg ions from B-sites can also migrate to A-sites during heat treatment process [84] and this migration may vary composition to composition. As a result, some of $Fe^{3+}$ ions migrate from A-site to B-site lead to an increase in magnetic moment of B-sites resulting a higher saturation magnetization for $x = 0.03$ in comparison to $x = 0.01$ & 0.02 and saturation magnetization is higher for $x = 0.04$ than that of $x = 0.05$.

## 3.6 Study of permeability

The frequency dependence of real part ($\mu_i'$) and imaginary part ($\mu_i''$) of complex permeability of $Mg_{0.5}Zn_{0.5}Y_xFe_{2-x}O_4$ ($0 \leq x \leq 0.05$) in the frequency range from 20 Hz to 120 MHz is presented in Fig. 10 (a and b). The $\mu_i'$ measures the ability of the materials to store energy, used to express the in phase relationship between the component of magnetic induction (B) and the alternating magnetic field (H) while the $\mu_i''$ measures the dissipation of energy in the materials and also used to express the component of B 90° out of phase with H. The formalism used to calculate the $\mu_i'$ and $\mu_i''$ of the $\mu_i^*$ are the following relations: $\mu_i' = L_s/L_0$ and $\mu_i'' = \mu_i' \tan\delta$, where $L_s$ is the self-inductance of the sample core and $L_0 = \frac{\mu_0 N^2 S}{\pi \bar{d}}$ is derived geometrically. Here, $L_o$ is the inductance of the winding coil without the sample core, N is the number of turns of the coil (N = 5), S is the area of cross section of the toroidal sample as given below: $S = d \times h$ and $d = \frac{d_2 - d_1}{2}$, where $d_1$ = inner diameter, $d_2$ = outer diameter and h= height and also $\bar{d}$ is the mean diameter of the toroidal sample as given below: $\bar{d} = \frac{d_2 + d_1}{2}$. The $\mu_i'$ remains fairly constant up to certain frequency, after which it increases to a maximum and after then sharply decreased with frequency. The frequency region in which the real part remains constant is defined as the utility zone of ferrites and demonstrated the suitability of the prepared ferrites in the wide range of frequency such as broadband pulse transformer and wide band read-write heads for video recording [39, 85]. On the other hand, the $\mu_i''$ decreases rapidly at lower frequency side and displays a characteristics peak at a higher frequency side where the $\mu_i'$ starts to decrease. This phenomenon is termed as the ferrimagnetic resonance [86]. Fig. 10 (d) shows the variation of permeability with Y contents (x). It is cleared from Fig. 10 [(a) and (c)] that the compositions exhibit the resonance at lower frequency due to their higher permeability and vice versa in accordance with the Snoek's limit $f_r \mu_i'$ = constant [87].

The values of $\mu_i'$ at 20 Hz frequency for different Y contents are shown in Fig. 10 (d). The variation of initial permeability of $Mg_{0.5}Zn_{0.5}Y_xFe_{2-x}O_4$ ($0 \leq x \leq 0.05$) can be explained by the following relation [88]: $\mu_i' \propto \frac{M_s^2 D}{\sqrt{K_1}}$ where $\mu_i$ is the initial permeability, $M_s$ the saturation magnetization, D the average grain size and $K_1$ the magneto-crystalline anisotropy constant. The value of $\mu_i'$ for x = 0.01 and 0.02 is greater than pure (x = 0.00) Mg-Zn ferrite. As $\mu_i'$ proportional to the square of the $M_s$ and directly proportional to the D, the $M_s$ is lowered for x = 0.01 and 0.02 but the value of D is increased from 1.22 µm to 3.65 µm and 3.75 µm for 0.01 and 0.02, respectively. The combine effect of both $M_s$ and D may remain the higher $\mu_i'$ values for $x$ = 0.01 and 0.02 than for $x$ = 0.00. After that, the values of $\mu_i'$ are found to decrease gradually for $x \geq$ 0.03. Although, The $M_s$ is slightly increased for $x$ = 0.03 than $x$ = 0.02, but the grains size decreases to 1.28 µm, resulting a decrease in $\mu_i'$ for $x$ = 0.03.

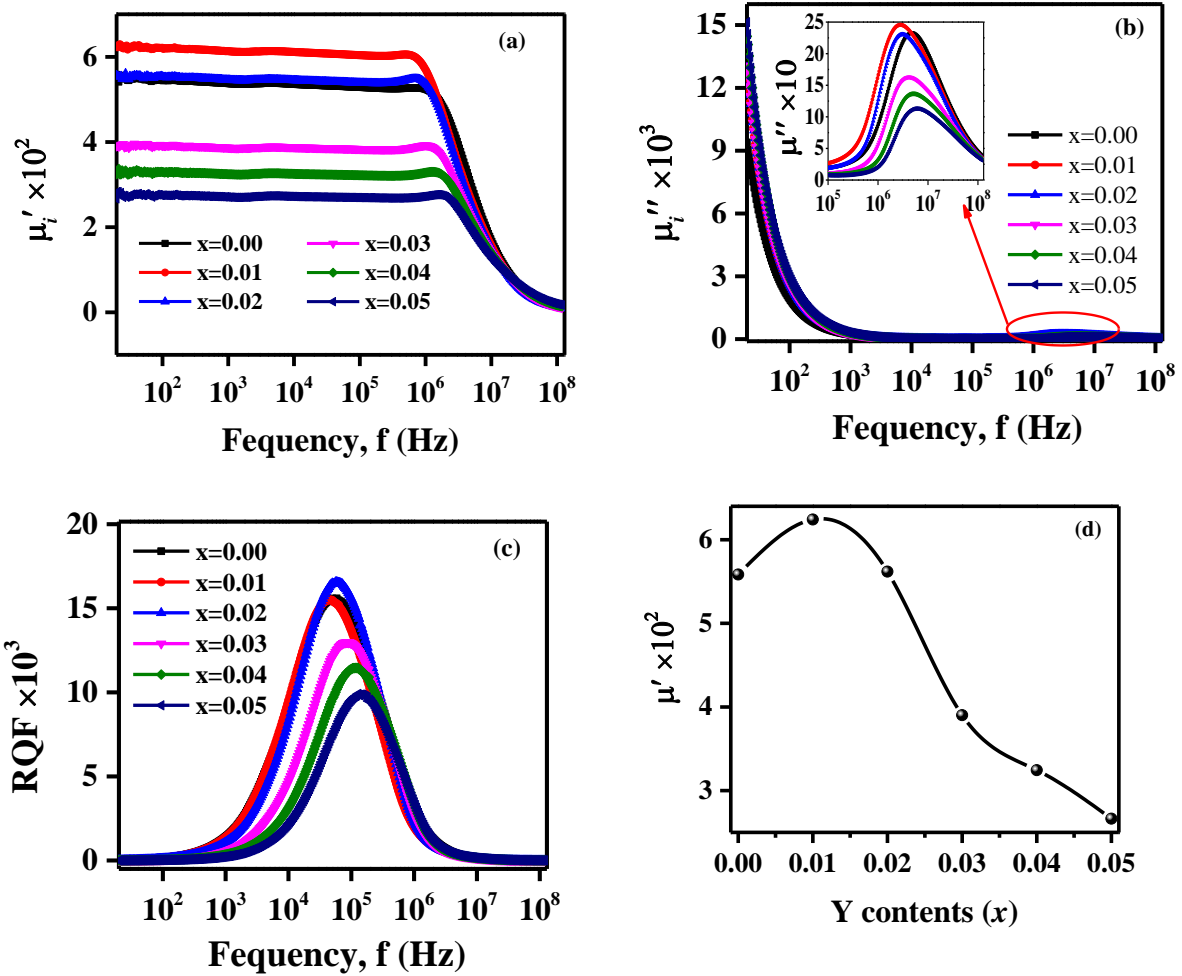

**Fig. 10** The frequency dependence of permeability (a) real part, (b) imaginary part, (c) relative quality factor and (d) the variation of $\mu_i$ with Y contents ($x$) of $Mg_{0.5}Zn_{0.5}Y_xFe_{2-x}O_4$ ($0 \leq x \leq 0.05$) at 20 Hz.

The frequency dependence of relative quality factor (RQF) of $Mg_{0.5}Zn_{0.5}Y_xFe_{2-x}O_4$ ($0 \leq x \leq 0.05$) ferrites is shown in Fig. 10 (c). The quality factor measures the performance of materials for used in filter application. The *RQF* is very low at very low frequency, found to increase with frequency, showing a characteristics peak, and decrease to low values at high frequency. The values of RQF go down beyond $10^6$ Hz where the loss factor [Fig. 10 (b)] starts to increase rapidly. The loss is resultant from the lagging of domain wall motion regarding the applied alternating magnetic field which is due to various domain defects [89] such as non-uniform and non-repetitive motion of domain-wall, bending over of domain-wall, localized variation of flux density and nucleation and annihilation of domain walls. The frequency for maximum in *RQF* shifts to higher frequency with Y contents except for $x = 0.01$, which is shifted slightly to lower frequency. The highest value of *RQF* is obtained for $x = 0.02$.

## 4. Conclusions

$Mg_{0.5}Zn_{0.5}Y_xFe_{2-x}O_4$ ($0 \leq x \leq 0.05$) ferrites have been successfully synthesized by standard ceramic technique. The single phase of spinel structure up to $x = 0.03$ and a secondary phase of $YFeO_3$ for $x \geq 0.04$ has been confirmed. An increase in lattice constant (both $a_{expt.}$ and $a_{theo.}$) is found ($\leq x = 0.03$ for experimental and thereafter decreases for $x \geq 0.04$, due to the secondary phase) for the substitution of Y ions with larger radius than that of Fe. The bulk density increases $x \leq 0.02$ from 4.006 g/cm$^3$ to 4.197 g/cm$^3$, there after it decreases with Y contents to 4.084 g/cm$^3$ for $x = 0.05$. On the contrary, the porosity decreases with Y contents $x \leq 0.02$, then increases with Y contents. FESEM images confirmed the homogeneous grain distribution and clear grain boundaries. Absence of any unwanted element is also confirmed from the EDS data. Y substitution has significantly influenced the electrical resistivity and dielectric constant. The ac electrical resistivity increases from $10^5$ to $10^6$ Ω-cm (at 20 Hz) for $x \geq 0.02$-0.04. A fairly good inverse relation between electrical resistivity and dielectric constant is observed for all the compositions. Existence of dielectric relaxation behavior is confirmed and the relaxation time is in the range between 15-31 nano seconds. The hysteresis loops confirm the soft ferromagnetic nature of the samples. The saturation magnetization is found to decrease (non-linearly) due to Y substitution. The $M_s$ is found to decrease up to $x = 0.02$ from 63.7 emu/g to 50.1 emu/g, thereafter it increases for $x = 0.03$ (58.3 emu/g) and decreases for $x = 0.04$ (40.1 emu/g) and finally increased again for $x = 0.05$ (46.8 emu/g). The initial permeability with wide stability regions ($\leq 10$ MHz) is found to increase for $x = 0.01$ and decreases gradually for higher Y contents. The enhanced electrical resistivity, reduced dielectric constants and appropriate coercive field make the Y-substituted Mg-Zn ferrites more suitable for high frequency device application.

## Acknowledgments

The authors are grateful to the Directorate of Research and Extension, Chittagong University of Engineering and Technology, Chattogram 4349, Bangladesh under grant number CUET DRE (CUET/DRE/2016-2017/PHY/005) for arranging the financial support for this work. We are thankful for the laboratory support of the Materials Science Division, Atomic Energy Center, Dhaka 1000, Bangladesh.